# A Comprehensive Study of Data Augmentation Strategies for Prostate Cancer Detection in Diffusion-weighted MRI using Convolutional Neural Networks

Ruqian Hao[1,2,3], Khashayar Namdar[3,4], Lin Liu[1], Masoom A. Haider[4,5,6], Farzad Khalvati[2,3,4]

[1] School of Optoelectronic Science and Engineering, University of Electronic Science and Technology of China, Chengdu, Sichuan, China

[2] Institute of Medical Science, University of Toronto, Toronto, ON, Canada

[3] The Hospital for Sick Children (SickKids), Toronto, ON, Canada

[4] Joint Department of Medical Imaging, Sinai Health System, University Health Network, University of Toronto, Toronto, ON, Canada

[5] Lunenfeld-Tanenbaum Research Institute, Sinai Health System, Toronto, ON, Canada

[6] Sunnybrook Research Institute, Toronto, ON, Canada

## Abstract

Data augmentation refers to a group of techniques whose goal is to battle limited amount of available data to improve model generalization and push sample distribution toward the true distribution. While different augmentation strategies and their combinations have been investigated for various computer vision tasks in the context of deep learning, a specific work in the domain of medical imaging is rare and to the best of our knowledge, there has been no dedicated work on exploring the effects of various augmentation methods on the performance of deep learning models in prostate cancer detection. In this work, we have statically applied five most frequently used augmentation techniques (random rotation, horizontal flip, vertical flip, random crop, and translation) to prostate Diffusion-weighted Magnetic Resonance Imaging training dataset of 217 patients separately and evaluated the effect of each method on the accuracy of prostate cancer detection. The augmentation algorithms were applied independently to each data channel and a shallow as well as a deep Convolutional Neural Network (CNN) were trained on the five augmented sets separately. We used Area Under Receiver Operating Characteristic (ROC) curve (AUC) to evaluate the performance of the trained CNNs on a separate test set of 95 patients, using a validation set of 102 patients for finetuning. The shallow network outperformed the deep network with the best 2D slice-based AUC of 0.85 obtained by the rotation method.

## 1- Introduction

Prostate cancer is the second most common type of cancer among men. Traditional screening methods such as Prostate Specific Antigen test and Digital Rectal Examination are usually at a high risk of low accuracy and overdiagnosis [1]. Wide application of Diffusion-Weighted Magnetic Resonance Imaging (DW-MRI) for prostate cancer detection has resulted in a higher sensitivity of predictive models. However, specificity is still low which is caused by dependence of examination results to experience and preference of the radiologist. Unnecessary biopsies are major consequence of the ongoing practice [2]. The recent advent of Convolutional Neural Networks (CNNs) has resulted in more focus on using Machine Learning (ML) for classification and diagnosis of prostate cancer. Deep learning algorithms which relies on CNNs with more layers, require a large amount of data. Nonetheless, it is difficult to get access to significant amounts of prostate cancer data due to the limited available data and existing regulations and protocols of medical images [3].

In order to accumulate enough data and improve performance of deep learning models for medical imaging classification tasks, multiple data augmentation strategies have been proposed. Cao *et al.* performed intensity normalization and basic augmentation methods (translate, scale, and flip) on a prostate Multi-parametric MRI (mp-MRI) dataset and then trained a CNN to detect prostate lesions. As a result, sensitivity improved by 9.8% [4]. Campanella *et al.* used a prostate core biopsy dataset consisting of 24,859 slides to classify prostate cancer. A ResNet34 model was trained without and with augmentation on the fly which consisted 90° rotations, horizontal flips, and color jitter. Nonetheless, the experiment results showed that data augmentation was not effective on the large-scale dataset. The highest balanced accuracy, which is the accuracy scaled by size of each individual class, on the model trained without augmentation was 0.95% higher than with augmentation [5].

Using data augmentation methods in the field of Medical Imaging is not quite novel, however, most studies have not quantified the effects of these techniques. Ishioka *et al.* utilized a series of transformations which included cropping prostate area of 261×261 pixels randomly and shaping them into parallelogram, and then generated a total of 2 million prostate MR images training dataset to train CNNs for prostate cancer detection. The experiments results showed that the best validation AUC increased from below 0.4 to 0.793 as the number of augmented training images increased from 0 to 2 million, and AUC value of this model for test dataset was 0.636 [6]. Wang *et al.* cropped each training T2-weighted prostate image of 360×360 into multiple sub-images of 288×288 pixels randomly. Deep convolutional neural network (DCNN) was trained for the automatic classification of prostate cancer, and the AUC result was 0.84 [7]. Liu *et al.* sliced 3D multiparametric MRI data at 7 different orientations and performed in-plane rotation, random shearing and one pixel translation of the lesions for each slice. At the end, 207,144 training samples were prepared. When training their proposed XmasNet, random mirroring was also performed on the fly. The same augmentation procedure was also applied to the validation and test sets. The AUC result was 0.84 for slices with biopsy in the augmented test set [8]. Mehrtash *et al.* used flipping and translation to generate 5-fold cross-validation prostate mpMRI datasets with 10,000 training and 2,000 validation samples for each fold, and then they trained a CNN for prostate cancer classification. AUC on the test set was 0.80 [9]. Esteva *et al.* rotated each skin cancer lesions image randomly between 0° and 359°. The largest upright inscribed rectangle was then cropped from the image and was flipped vertically with a probability of 0.5. As a result, the number of samples in training data size was enlarged by a factor of 720. Then a CNN was trained to classify skin cancer and the ultimate AUC reached over 91% [10].

In the natural image domain, multiple augmentation methods have been investigated to improve recognition and classification accuracy. Ding *et al.* compared three types of data augmentation operation including translation, speckle noising (pointwise multiplying the mean filtered SAR image by random samples from exponential distribution), and pose synthesis (rotate a SAR image and combine it with the original one linearly), and the experiment results showed combining all types of augmentation operations is a practical approach for target recognition in challenging conditions of target translation, random speckle noise, and missing pose [11]. Lv *et al.* proposed five data augmentation methods to face images including landmark perturbation (using landmark to do translation, rotation, shearing, and scaling) and four synthesis methods (hairstyles, glasses, poses, and illuminations), and trained a CNN with a dataset by concatenating all data augmentation features to recognise face. They achieved an accuracy of 94.08%, reducing the error by 26% compared with the one without data augmentation [12].

In the previous studies in prostate cancer detection using deep learning methods [4]–[9], as a conventional method for expanding the size of dataset, a data augmentation method or a combination of different methods they were randomly picked. While some augmentation strategies and their combinations have been investigated for various medical imaging tasks, there has been no dedicated work on exploring effects of different augmentation methods on performance of deep learning models in prostate cancer detection. Our work focusses on studying the effects of several conventional data augmentation methods on the performance of a shallow and deep CNN, which can fill the gaps in this research field and provide guidance on what data augmentation methods should be employed with CNNs of different depths in future prostate cancer detection research.

In this work, we have statically applied five most popular augmentation techniques (i.e. random rotation, horizontal flip, vertical flip, random crop, and translation) to the prostate DW-MRI training dataset of 217 patients separately and trained a shallow as well as a deep 2D CNN on the five augmented sets respectively. Finally, we used AUC which is a commonly used metric for binary classification of unbalanced medical image datasets [13], to evaluate performance of the trained CNNs of different depths on a separate test set of 95 patients, using a validation set of 102 patients for finetuning. The classification was done on 2D DW-MRI slices and the highest AUC on the test set was 85.04% achieved by training the shallow CNN on the augmented dataset using random rotation method. To gain a deeper insight into augmentation methods, we used a CNN heatmap generator to investigate the best way to augment images in different DW-MRI sequences; whether to treat images of all sequences as a single unit and augment as such or apply the augmentation methods to images of each sequence independently.

Our proposed method is fully automated with no user intervention. The 2D slices along with their labels (cancer or no cancer) were fed to the CNNs for training. In test phase, the 2D slices are fed into the CNN regardless whether they contained prostate and hence, eliminating the need to manually (or automatically) segment prostate. The fully automated nature of our algorithm makes it a better candidate for clinical integration of prostate cancer management. It can process a large set of prostate DW-MRI image and produce results in minutes.

## 2- Methods

In this section, we present dataset description, different augmentation methods that we have used for the classification of prostate cancer in DW-MRI dataset, augmentation process for each DW-MRI sequences, an off-the-shelf deep CNN and our proposed shallow CNN architectures, and the GradCAM attention network [14] we used to evaluate robustness of the results.

### 2-1- Dataset

The dataset was obtained as part of a retrospective single institution study and the institutional review board approval was received. All patients had a positive MRI. All exams were performed on a 3T MRI system without an endorectal coil MRI. Our proprietary dataset included DW-MRI images of 414 patients, corresponding to 10,128 2D 6-channel slices. Each slice consisted of a DWI sequence (i.e. 6-channels in the domain of image processing) comprising of an ADC map and five b-values (0, 100, 400, 1000, and 1600 $s\ mm^{-2}$). The dataset was split into training set (217 patients 5,300 2D slices), validation set (102 patients, 2,500 2D slices), and test cohort (95 patients, 2,328 2D slices). Every slice was labelled according to the Gleason Score from targeted biopsy results. When the Gleason score was larger than 6, the case was considered as clinically significant prostate cancer and the slice was labeled as positive. 144×144 pixels was our fixed image size across the whole dataset, and there were 6 input channels which were stacked in the order of ADC, b0, b1000, b100, b400, and b1600.

### 2-2- Augmentation methods

Augmentation methods in deep learning can be applied either statically or on the fly. Static data augmentation refers to appending the augmented data to the training dataset and using the augmented dataset for training the model. On the other hand, when we randomly augment the data in each batch while the model is being trained, it is, in fact, on the fly data augmentation applied. Although on the fly augmentation is more efficient in terms of computational and storage resources, static approach provides more flexibility to data augmentation research. For example, with static augmentation, we can go back and manually check each augmented example. Therefore, we chose to implement our algorithms statically.

To reduce bias, achieve better generalization, and compensate for a relatively small amount of training dataset, random rotation, horizontal flip, vertical flip, random crop, and translation are the most frequently used augmentation techniques in medical imaging tasks. For each image $I(x,y)$ in the training set, we utilize an augmentation strategy to obtain transformed image $I(x',y')$.

2.2.1. Random rotation

Our random rotation augmentation acts clockwise by an angle selected from the range of degrees (-degrees, +degrees) randomly. Pixels outside the rotated area will be filled with 0. The rotation formula is given by Equation (1) and Equation (2).

$$\begin{bmatrix} x' \\ y' \\ 1 \end{bmatrix} = \begin{bmatrix} cos\,\theta & sin\,\theta & 0 \\ -sin\,\theta & cos\,\theta & 0 \\ 0 & 0 & 1 \end{bmatrix} * \begin{bmatrix} x \\ y \\ 1 \end{bmatrix} \tag{1}$$

$$x' = x * cos\,\theta - y * sin\,\theta, \; y' = x * cos\,\theta + y * sin\,\theta \tag{2}$$

## 2.2.2. Horizontal flip

The horizontal flip augmentation flips the input Image along its vertical (left to right) axis randomly with a given probability. The horizontal flip formula is given by Equation (3) and Equation (4).

$$\begin{bmatrix} x' \\ y' \\ 1 \end{bmatrix} = \begin{bmatrix} -1 & 0 & 0 \\ 0 & 1 & 0 \\ 0 & 0 & 1 \end{bmatrix} * \begin{bmatrix} x \\ y \\ 1 \end{bmatrix} \tag{3}$$

$$x' = -x, \; y' = y \tag{4}$$

## 2.2.3. Vertical flip

The vertical flip augmentation flips the input image along its horizontal (top to bottom) axis randomly with a given probability. The vertical flip formula is given by Equation (5) and Equation (6).

$$\begin{bmatrix} x' \\ y' \\ 1 \end{bmatrix} = \begin{bmatrix} 1 & 0 & 0 \\ 0 & -1 & 0 \\ 0 & 0 & 1 \end{bmatrix} * \begin{bmatrix} x \\ y \\ 1 \end{bmatrix} \tag{5}$$

$$x' = x, \; y' = -y \tag{6}$$

## 2.2.4. Random crop

The sub-image is a square and the side length of the square (i.e. crop size) is given in advance which is smaller than the minimum side length of the original image. The sub-image is included in the original image, and its center position is randomly selected. Using this method, our random crop augmentation crops the original image into given size sub-image randomly. Then the cropped images are scaled up to the original image size by an upsampling technique called Nearest-neighbor interpolation [15].

## 2.2.5. Translation

The translation augmentation shifts the input image in horizontal and vertical directions with a given maximum absolute fraction. For the example of translate fraction = (a, b) on an image of width W and height H, the horizontal shift (dx) and vertical shift (dy) are randomly selected from the uniform distributions over ranges [-W⋆a , W⋆a] and [-H⋆b , H⋆b] respectively. The translation formula is given by Equation (7) and Equation (8).

$$\begin{bmatrix} x' \\ y' \\ 1 \end{bmatrix} = \begin{bmatrix} 1 & 0 & dx \\ 0 & 1 & dy \\ 0 & 0 & 1 \end{bmatrix} * \begin{bmatrix} x \\ y \\ 1 \end{bmatrix} \tag{7}$$

$$x' = x + dx, \ y' = y + dy \tag{8}$$

For a more intuitive display effect, we have applied these five augmentation methods on an Apparent Diffusion Coefficient (ADC) image of prostate cancer. Augmented images are shown in Figure 1.

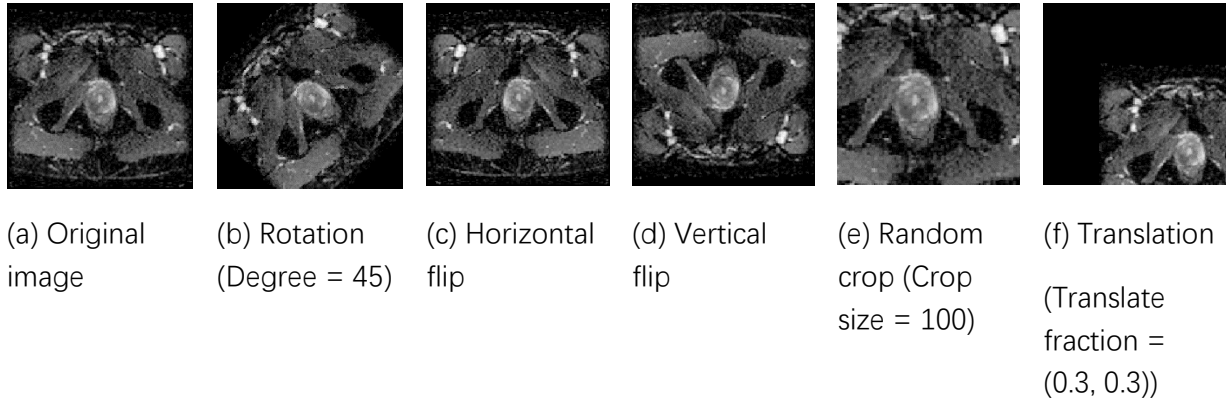

(a) Original image

(b) Rotation (Degree = 45)

(c) Horizontal flip

(d) Vertical flip

(e) Random crop (Crop size = 100)

(f) Translation (Translate fraction = (0.3, 0.3))

Figure 1. Effects of the 5 augmentation methods applied to a prostate ADC image

### 2-3- Independent Channel Augmentation Process

In the field of natural image augmentation, augmentation methods are usually applied to all three Red/Green/Blue (RGB) channels at the same time. While the RGB channels represent different color components of a natural image, the concept of channel in medical imaging is different. In this work, each input image consists of 6 channels (sequences): ADC, b0, b1000, b100, b400, and b1600 as shown in Figure 2. These channels indicate the amount of water diffusion in prostate gland and have different characteristics and features which can be used to distinguish between normal tissue and tumor region [16].

To implement our augmentation algorithms, two different options were considered. We could either apply our augmentation algorithm to each channel independently or stack channels together in order to have all channels manipulated similarly. Intuitively, the independent approach imposes more variance to the data which translates to a richer data. It should be noted that the same reasoning does not apply to natural images with RGB channels as all three channels measure texture in a similar way. We evaluated the two scenarios for our best augmentation setting. As it is depicted in Figures 2 to 4, in one trial, all the 6 channels image were rotated as a unit identity at a rotation degree from range of [-50, 50] and in other trial, they were rotated independently. Our experiment showed that when each channel is augmented separately, the test AUC result improved by 1.75%.

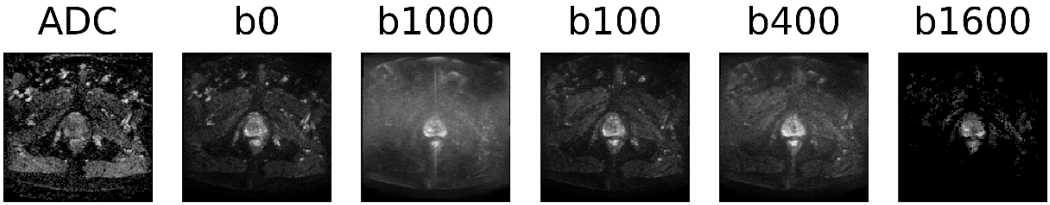

Figure 2. Original image of 6 channels

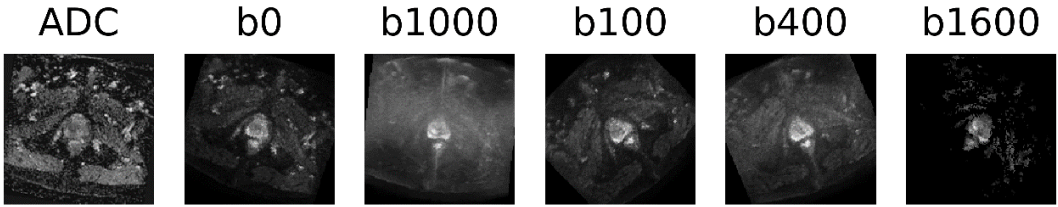

Figure 3. Applying random rotation in each channel independently

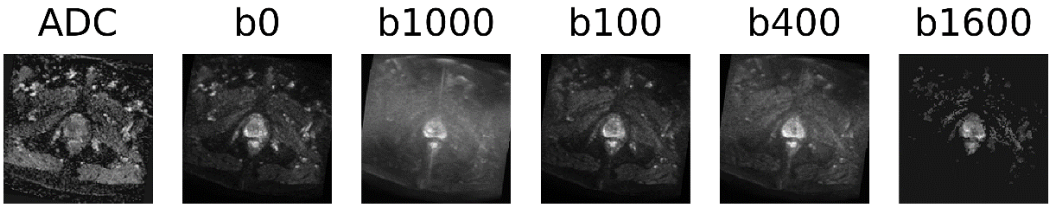

Figure 4. Applying random rotation in all channels at the same time

### 2-4- Gradient-weighted Class Activation Mapping (Grad-CAM)

In addition to AUC results for different augmentation method, we used a CNN heatmap generator to investigate whether treating all images of six channels as a single unit in applying augmentation performs better than when each channel is augmented independently. The hypothesis for using CNN visualization was that If the original sample is misclassified and augmentation helped to classify it correctly, this will be reflected in the heatmap. To do so, we input the original and augmented images to the trained CNN and obtain class predictions, and then computed Grad-CAM visualizations for each of the predicted classes. Proposed by R. Selvaraju *et al.*, Grad-CAM determines which regions of input are more important for predictions from CNN models. Grad-CAM uses the class-specific gradient information flowing into the final convolutional layer of a CNN to generate a coarse heatmap which indicates the important regions in the image [14].

### 2-5- CNN architecture

To investigate the effect of network depth on optimal augmentation settings, we used two different CNN architectures for our experiments: a shallow CNN [17] and a deep CNN whose design is inspired by VGGNet [18]. The shallow CNN consists of 4 convolutional layers, 3 max-pooling layers, 3 dropout layers, and 3 fully connected layers. The deep CNN is composed of 10 convolutional layers, 4 max-pooling layers and 4 fully connected layers. Architectures of the shallow and deep CNN are shown in Figure 5 and Figure 6, respectively. Configurations and output size of each layer in the shallow and deep CNN are listed in Table 1 and Table 2, respectively.

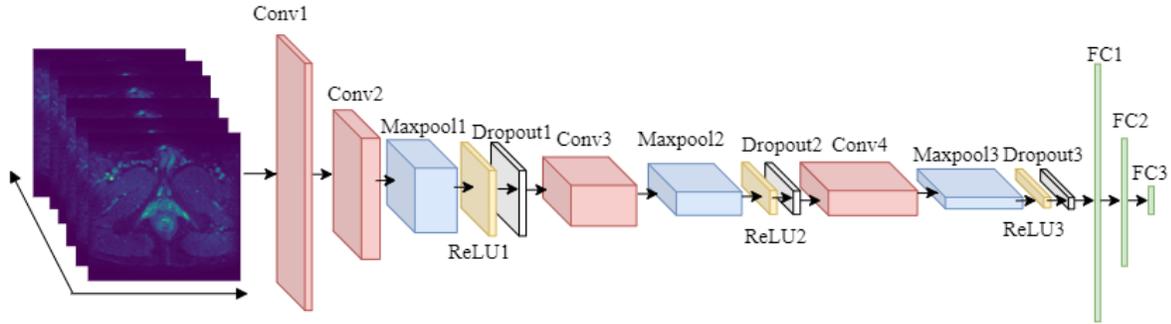

Figure 5. Architecture of the shallow CNN

Table 1. Configuration of the shallow CNN

| Layer | Kernel size | Stride | padding | probability | Output size |
|---|---|---|---|---|---|
| Conv1 | 3×3 | 1 | 0 | | 128×142×142 |
| Conv2 | 3×3 | 2 | 0 | | 256×64×64 |
| Maxpool1 | 2×2 | 2 | | | 256×32×32 |
| Dropout1 | | | | 0.1 | |
| Conv3 | 3×3 | 2 | 0 | | 512×15×15 |
| Maxpool2 | 2×2 | 2 | | | 512×7×7 |
| Dropout2 | | | | 0.1 | |
| Conv4 | 3×3 | 2 | 0 | | 1024×3×3 |
| Maxpool3 | 2×2 | 2 | | | 1024×1×1 |
| Dropout3 | | | | 0.1 | |
| FC1 | | | | | 256×1 |
| FC2 | | | | | 64×1 |
| FC3 | | | | | 2×1 |

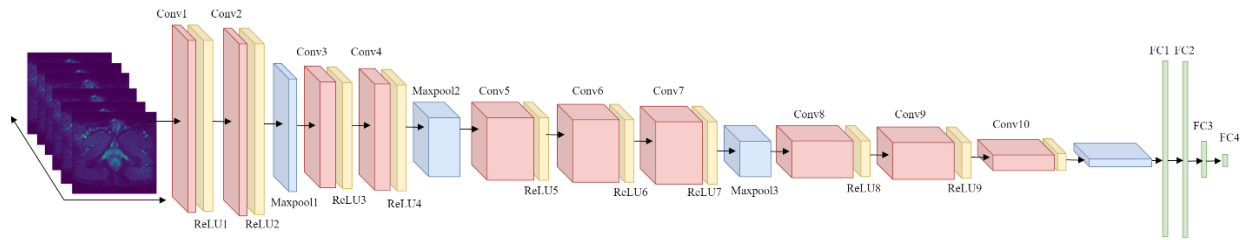

Figure 6. Architecture of the deep CNN

Table 2. Configuration of the deep CNN

| Layer | Kernel size | Stride | padding | Output size |
|---|---|---|---|---|
| Conv1 | 3×3 | 1 | 1 | 64×144×144 |
| Conv2 | 3×3 | 1 | 1 | 64×144×144 |
| Maxpool1 | 2×2 | 2 | | 64×72×72 |
| Conv3 | 3×3 | 1 | 1 | 128×72×72 |
| Conv4 | 3×3 | 1 | 1 | 128×72×72 |
| Maxpool2 | 2×2 | 2 | | 128×36×36 |
| Conv5 | 3×3 | 1 | 1 | 256×36×36 |
| Conv6 | 3×3 | 1 | 1 | 256×36×36 |
| Conv7 | 3×3 | 1 | 1 | 256×36×36 |
| Maxpool3 | 2×2 | 2 | | 256×18×18 |
| Conv8 | 3×3 | 1 | 1 | 512×18×18 |
| Conv9 | 3×3 | 1 | 1 | 512×18×18 |
| Conv10 | 3×3 | 2 | 1 | 512×9×9 |
| Maxpool4 | 2×2 | 2 | | 512×4×4 |
| FC1 | | | | 1024×1 |
| FC2 | | | | 1024×1 |
| FC3 | | | | 256×1 |
| FC4 | | | | 2×1 |

The CNNs proposed in this paper are both trained with Cross Entropy loss function and optimized by Stochastic Gradient Descent (SGD) with momentum [19]. Weights of layers were initialized by Xavier method [20] and the biases were initialized to random values picked from a uniform distribution over 0 to 1.

## 2-6- Settings of Training

In this work, we varied the augmentation hyperparameters across specified ranges. Each set of hyperparameters resulted in a new classification problem. Augmentation was only performed on the training set. To create the ultimate training data, the augmented slices were statically appended to the original training data. In extreme cases such as augmenting by zero degrees of rotation, this method results in simple oversampling.

Our workflow is shown in Figure 7, which consists of the following main steps: Firstly, we varied the augmentation hyperparameters across specified ranges for each of the 5 augmentations: random rotation, horizontal flip, vertical flip, random crop, and translation. In random rotation we applied a rotation angle ranging from 0 degree to 180 degrees, with interval of 5 degrees. In horizontal and vertical flip augmentations, we worked within the probability range 0 to 1, with 0.05 steps. Images in the training set was assigned random numbers as probability, which were used to decide whether or not flip the image. When applying random crop, we took the crop size from 70 to 140 pixels, with intervals of 5. In translation, the translate fraction was selected from 0.0 to 0.5, with 0.01 steps. It must be noted that augmentation was only performed on the training set.

Secondly, we used oversampling strategy by adding the augmented sets to the original training dataset. By performing this step, we produced five different augmented training datasets each with 10,600 slices.

Thirdly, after performing normalization across the entire dataset, we trained the shallow CNN and deep CNN independently on the five augmented training datasets. When training the shallow CNN, the learning rate was set to 0.001, batch size was set to 1, L2 regularization penalty was set to 0.001, and momentum in SGD was 0.8 [17]. The hypermeters in the deep CNN were almost the same except for the learning rate, we decreased the learning rate to avoid divergence. Finally, we calculated AUC as a measure to evaluate performance of the trained CNNs.

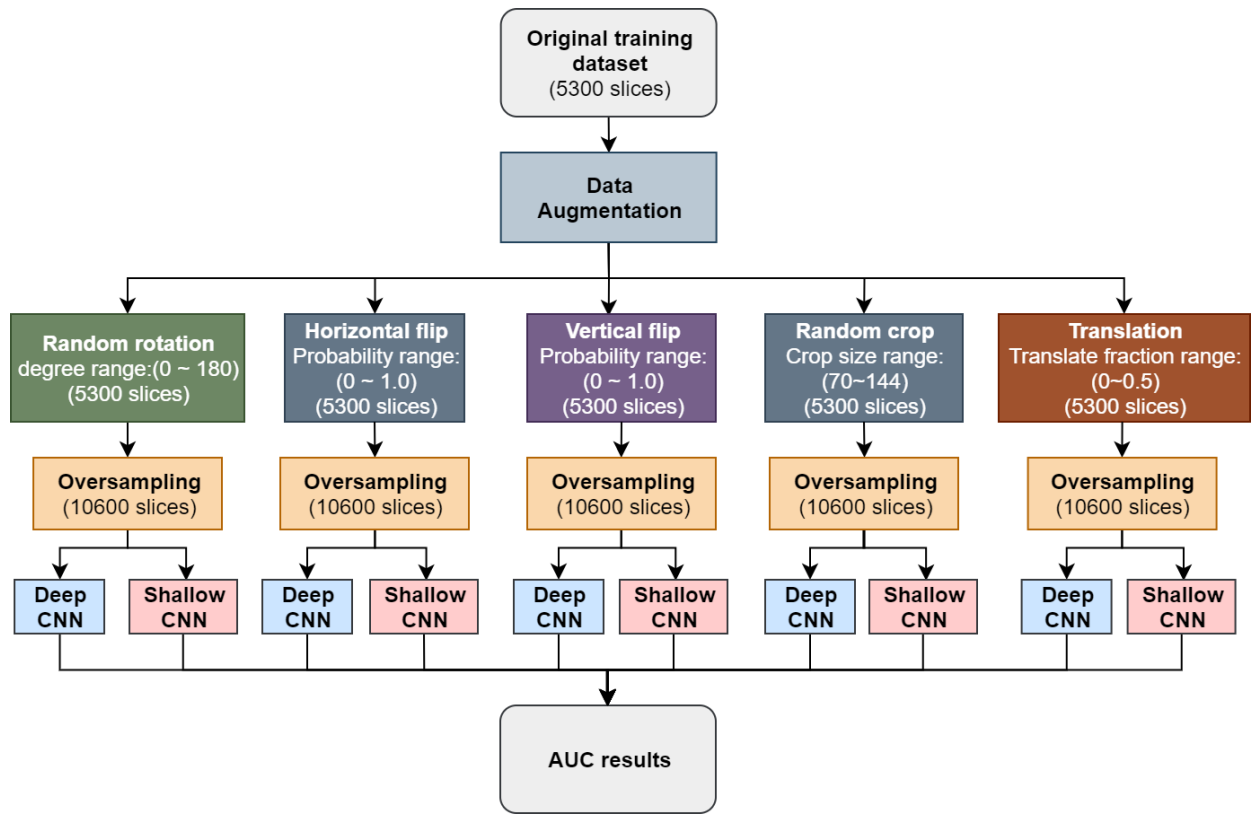

Figure 7. Workflow chart

## 3-  Results

### 3-1- The Original AUC and Baseline AUC

For the purpose of comparing the effect of different augmentation methods from the same starting point with minimal computational cost, all experiments were performed only once with a fixed random seed. The original AUC and baseline AUC are presented in Table 3. The original AUC was calculated without oversampling technique, which means CNNs were trained on the original training dataset. The baseline AUC was computed when there were no augmentation methods applied to the training dataset, except for duplicating the training set. As it is seen from Table 3, the AUC improvement was negligible or nonexistent when oversampling technique was applied to train the shallow and deep CNNs, respectively. All the experiments were conducted on a Nvidia DGX station platform, using Python 3.7.3 and PyTorch 1.2.0.

Table 3. The Original AUC and Baseline AUC

| AUC (%) | The original AUC | | The baseline AUC | |
|---|---|---|---|---|
| | Validation dataset | Test dataset | Validation dataset | Test dataset |
| The shallow CNN | 82.59 | 78.61 | 82.63 | 80.23 |
| The deep CNN | 80.23 | 74.81 | 80.23 | 74.81 |

### 3-2- AUC Results of Using Different Augmentation Methods

Figure 8 shows the visualizations of validation and test AUC results on shallow CNN and deep CNN when varying the augmentation hyperparameters across specified ranges for five augmentation methods (random rotation, horizontal flip, vertical flip, random crop, and translation). There are five rows and two columns in Figure 8, a total of ten subfigures. In each subfigure, the solid lines indicate the AUC values obtained by using the data augmentation method corresponding to the parameters of the horizontal axis, and the dotted lines represent the baseline AUC which was computed when there were no augmentation methods applied to the training dataset (training data was only duplicated). The two colors blue and red in each subfigure represents AUC results for the validation set and test set, respectively.

By comparing each row in Figure 8, it is seen how different depths of CNN architecture affected the AUC results when applying the same data augmentation method to the training dataset; overall, it can be observed that the shallow CNN performs better than the deep CNN. We can also compare the subfigures in each column, which presents the AUC results of training the same CNN with different data augmentation methods. The experiment results show that random rotation and translation were the most efficient augmentation methods when using the shallow CNN and the deep CNN, respectively.

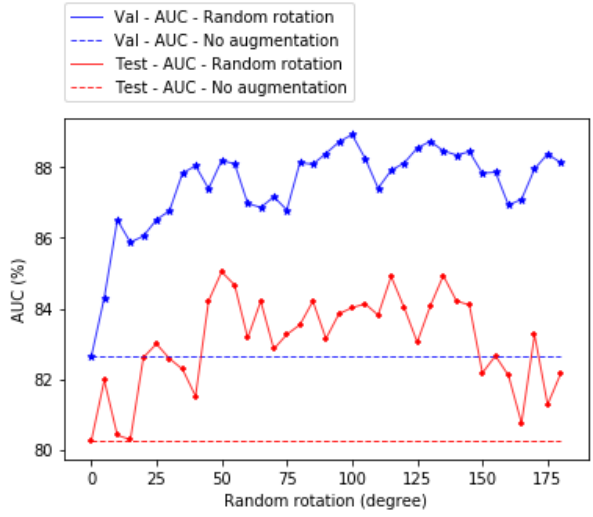

(a) Shallow CNN: Random Rotation

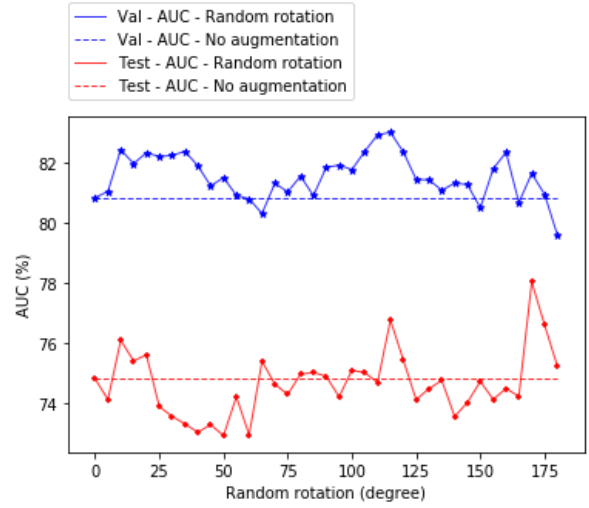

(b) Deep CNN: Random Rotation

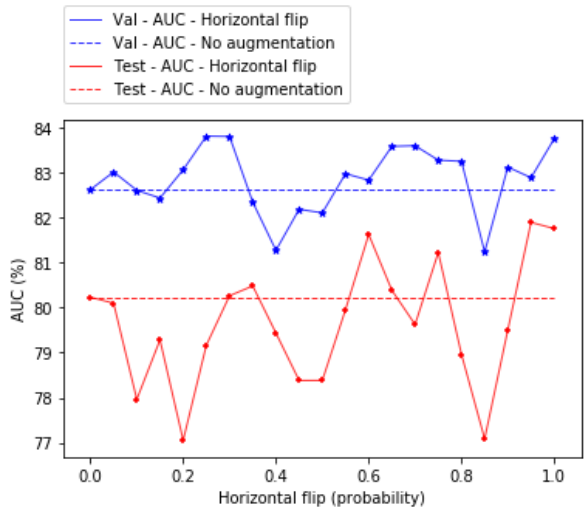

(c) Shallow CNN: Horizontal Flip

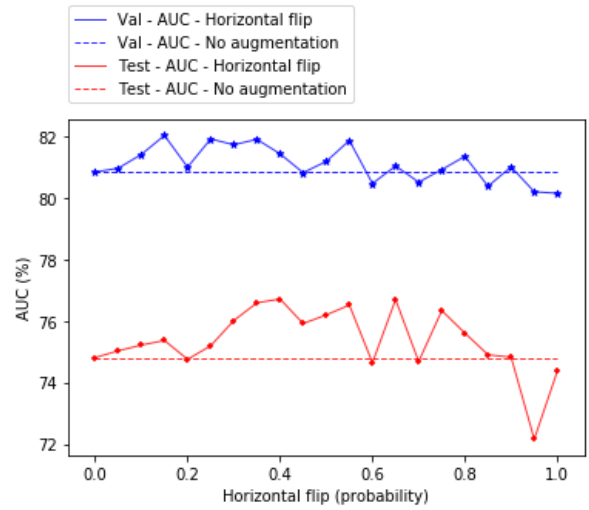

(d) Deep CNN: Horizontal Flip

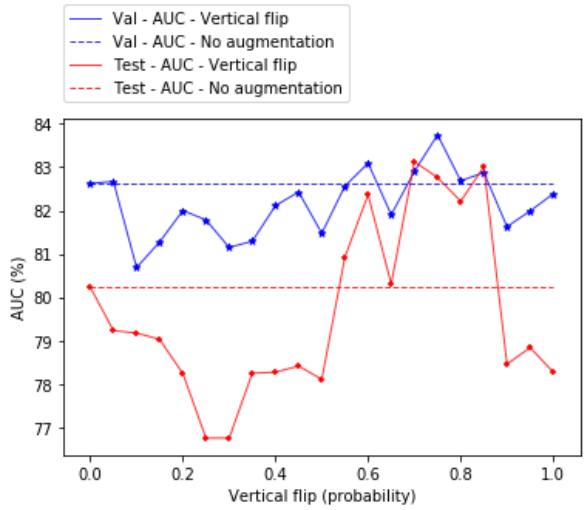

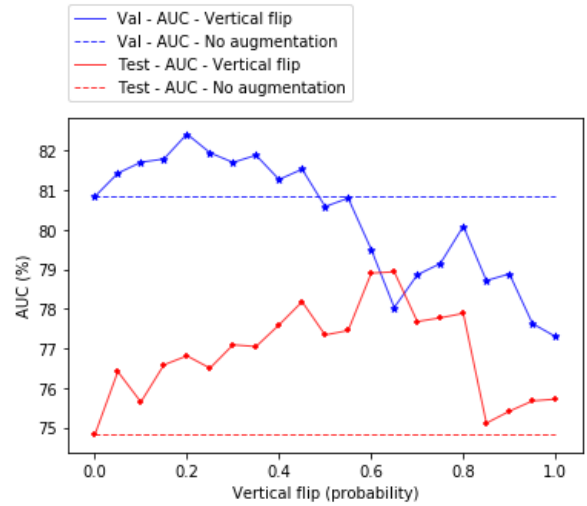

(e) Shallow CNN: Vertical Flip

(f) Deep CNN: Vertical Flip

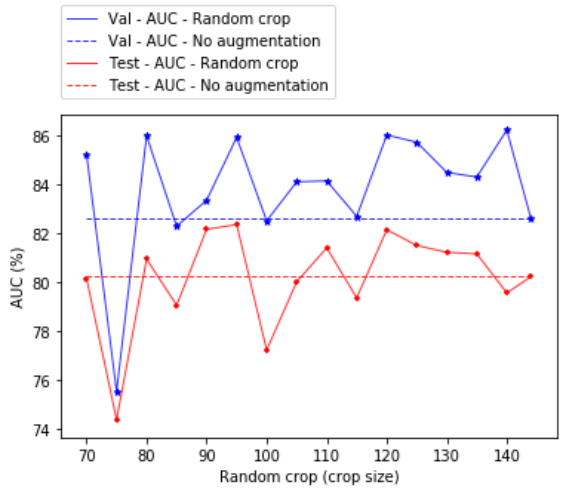

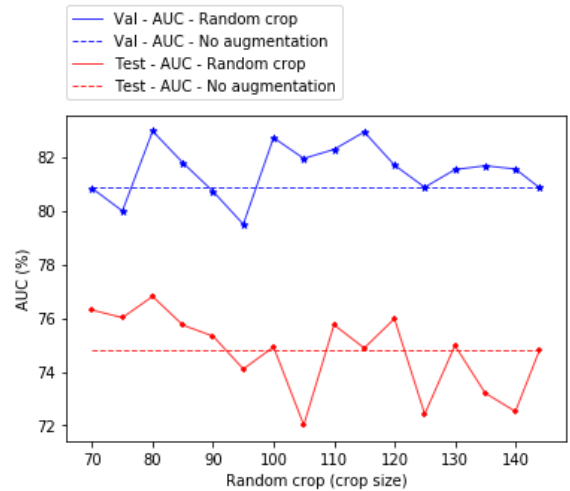

(g) Shallow CNN: Random Crop

(h) Deep CNN: Random Crop

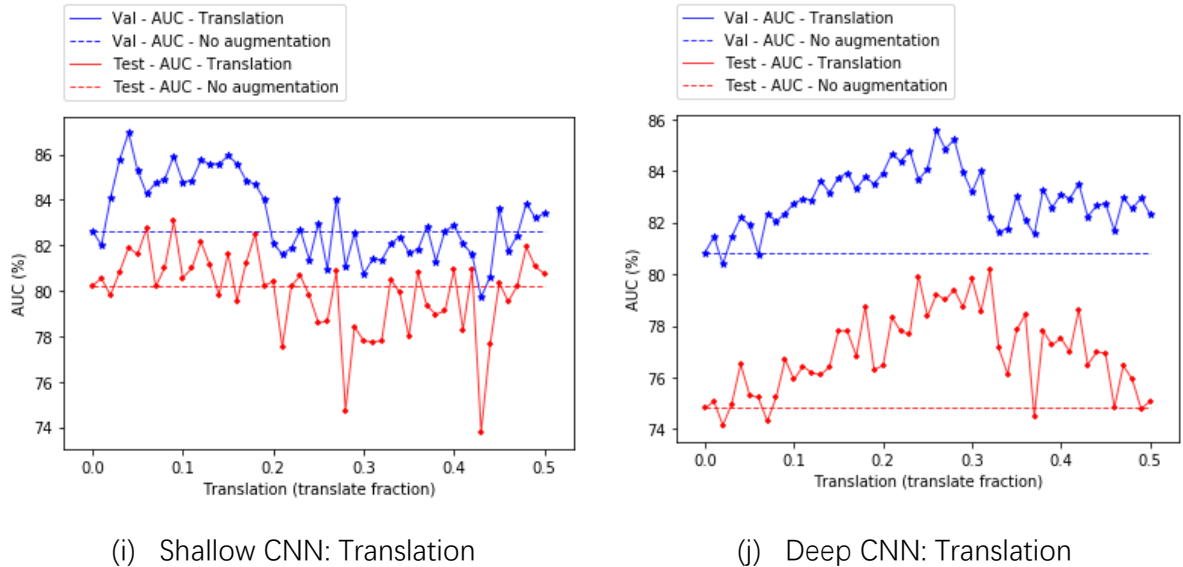

(i) Shallow CNN: Translation        (j) Deep CNN: Translation

Figure 8. Comparison of AUC results of using five augmentation methods to train shallow and deep CNNs

Table 4 lists the highest validation and test AUC results of both Shallow CNN and Deep CNNs for the five augmentation methods. As it can be seen, the highest validation AUC of 88.93% and test AUC of 85.04% were obtained when the random rotation technique (with [-100, 100] and [-50, 50] degree) was applied to train shallow CNN for prostate cancer classification.

Table 4. Comparison of the Highest AUC Results on Shallow and Deep CNNs

| The Highest AUC (%) | Augmentation Methods | | | | | | | | | |
|---|---|---|---|---|---|---|---|---|---|---|
| CNN Type | Random rotation | | Horizontal flip | | Vertical flip | | Random crop | | Translation | |
| | Val | Test | Val | Test | Val | Test | Val | Test | Val | Test |
| The shallow CNN | **88.93** | **85.04** | 83.81 | 81.90 | 83.74 | 83.12 | 86.23 | 82.37 | 86.95 | 83.06 |
| The Deep CNN | 83.01 | 78.04 | 82.03 | 76.72 | 82.43 | 78.94 | 82.95 | 76.80 | 85.56 | 80.16 |

Figure 9 visualizes the comparison of the test AUC results of Shallow CNN and Deep CNN. For each of the augmentation methods, it shows the highest AUC, the lowest AUC, and the average AUC in the test dataset among the range of hyperparameters for the augmentation methods (e.g., rotation degree).

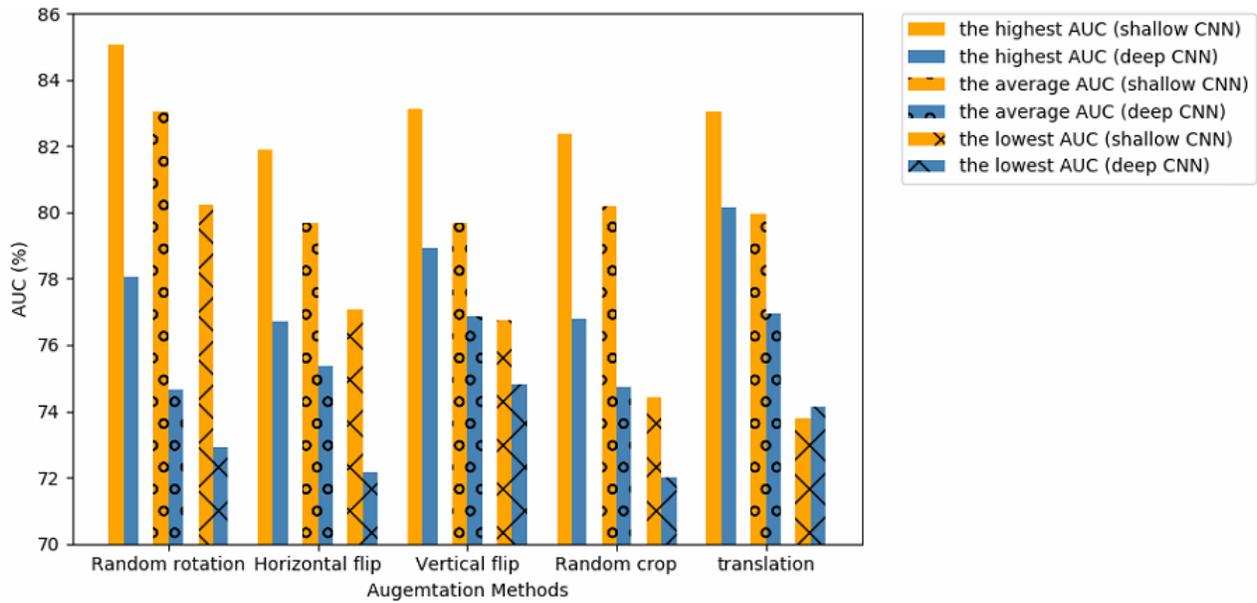

Figure 9. Comparison of Test AUC Results on Shallow and Deep CNNs

### 3-3- AUC Results on Different Devices

For the purpose of comparing the effect of the randomization, same augmentation method with different seedings on different devices was tested. For this purpose, Nvidia TITAN X (Pascal) GPU was used to run the code. It would be computationally prohibitive to try multiple seeds for each experiment. Hence, we used the augmented dataset using random rotation method to train the shallow CNN and compared the test AUC results with the results obtained from the previous run using the DGX platform. The comparison of test AUC results on different devices is visualised in Figure 6.

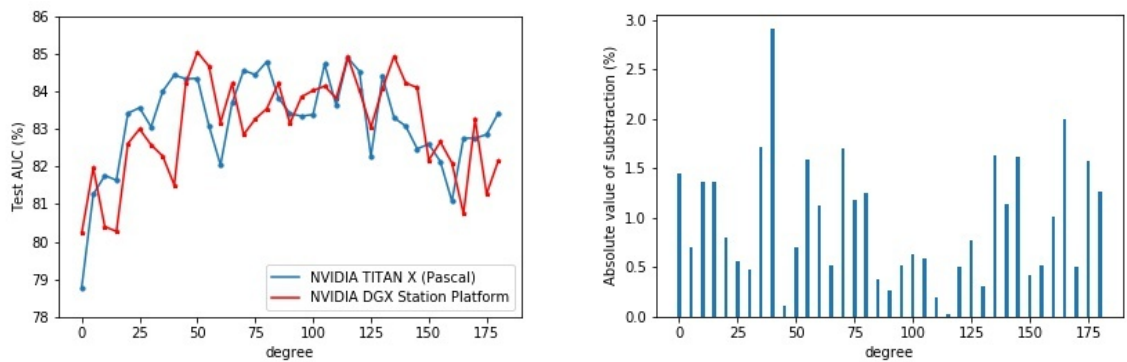

(a) Trend graph of the test AUC results      (b) Histogram of absolute values of AUC subtraction

Figure 10. Comparison of test AUC results on different devices.

### 3-4- Grad-CAM Visualizations of Samples and Corresponding Augmented Samples

In order to deepen our understanding of data augmentation process, we designed a test using a CNN heat map generator. From the training set, examples and their corresponding augmented instances were fed to the trained shallow CNN and their Grad-CAM visualizations were calculated. Although it was not practical to manually check every single case, through examining 50 random cases, it we realized that when the model is able to correctly classify an example, the heatmap correctly corresponds to the location of the prostate in the image. Otherwise, the heatmap points to an area outside prostate.

When the original sample is misclassified while the augmented sample is classified correctly, it can be observed from the visualization map that the augmentation led the network to detect correct information in the image. We used our GradCAM setting to provide more evidence to our earlier claim about independent channel augmentation in medical imaging. Shown in Figure 11, we randomly picked a misclassified positive case whose channel-independent augmented version was correctly classified and calculated their GradCAMs. Because it gives a better representation, only ADC channel is plotted in the figure. In the next step, we rotated the same image, however, not channel independent. As reflected in Figure 12, fixed rotation (treating all channels as a single unit) is not capable of correcting CNN attention and hence, the augmented sample is also misclassifed.

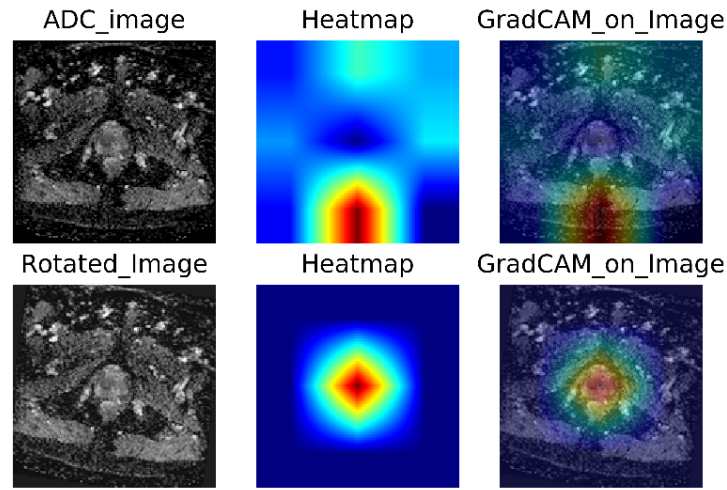

Figure 11. GradCAM visualizations of a misclassified positive example and its channel-independent augmentation pair

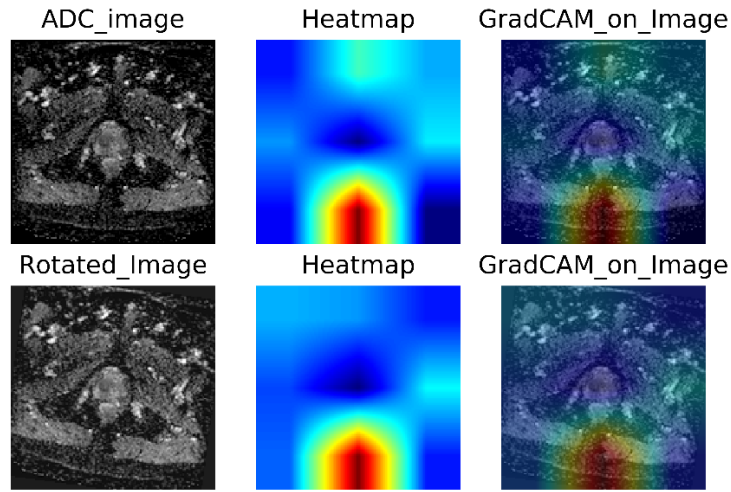

Figure 12. GradCAM visualizations of a misclassified positive example and its channel-depended augmentation pair

## 4- Discussion

Based on the presented results in Table 3 and Figure 8, random rotation is the most efficient augmentation method for prostate cancer detection when the CNN architecture was shallow. After training the shallow CNN on the augmented set using random rotation, we achieved the highest AUC of 85.04% on test set when degree range was set to (-50, 50), which was 5.2% higher than the baseline AUC (80.84%). As mentioned before, baseline refers to oversampling whose AUC is higher than the original dataset (i.e., not oversampling and no augmentation).

Translation is the most efficient augmentation method when the deep CNN was applied for prostate cancer classification. With the translate fraction set to (0.32, 0.32), the AUC on the test set was the highest (80.16%); 7.15% higher than the baseline AUC (74.81%).

As shown in Figure 9, except for the worst translation case, the shallow CNN performs better than the deep CNN. in other words, the effect of data augmentation was more prominent for the shallower architecture. The higher performance of the shallower network may be explained by the noisy nature of MRI images. Due to the environment, equipment and the performance of different doctors, MRI images usually contain a significant amount of noise [21]. We hypothesize that our proposed shallow CNN architecture is rich enough in terms of number of network weights, to not get stuck in the region of underfitting and light enough to avoid overfitting due to noise. However, the deep CNN architecture is prone to overfitting due to inheriting noise in MRI images.

For the shallow CNN, random rotation worked best because prostate is symmetrical. As a result, horizontal flip should not be able to add any value to the dataset other than a semi-oversampling effect. The nature of convolution operation in each operation minimizes effect of vertical flip on the other hand. The padding used in translation negatively affects the convolution operation [22]. Therefore, translation is not a good option for prostate, especially in cases where the tumor is small. Rotation is an effective augmentation method which creates asymmetric images. In the case of prostate cancer detection, asymmetry pushes the network to learn decisive patterns of the given image. For the deeper CNN, while random rotation is still promising, translation in some special configurations surpasses other methods. The reason may be because due to arbitrary noise reduction which helps the network to void overfitting.

It was interesting to observe that the results of augmentation methods on the validation and test sets are highly correlated. This indicates our validation and test cohorts are both good representatives of the original population. Furthermore, it can be inferred that our augmentation technique is effective for the true distribution and is not an arbitrary fit to our sample space.

A similar DW-MRI prostate cohort was used by Yoo *et al.* to develop a CNN-based pipeline using a modified ResNet architecture for the same prostate cancer detection [23]. The main differences between the two methods are as follows. First, [23] used a deep CNN (ResNet) while our best performance was achieved using a shallow CNN. Second, as opposed to our method, no augmentation was used in [23]. Finally, In [23], the slices that did not contain any portion of prostate gland were manually eliminated and the remaining slices were center-cropped with a fixed size. This requires manual markings of the apex and base slices of the prostate. In our work, however, detection of prostate cancer is fully automated with no preselection of slices. The consequence of our approach is reflected in our sample sizes where for example our test set contains 2,328 slices as compared to their 1,486 slices in [23]. Although our AUC result (85.04%) is slightly less than the average slice-level result reported in [23] (mean AUC of 86%), our method is fully automated with no need for any user intervention.

While it was not computationally feasible to repeat our results with different network initializations, we evaluated the results for the best configuration (rotation). Our parallel experiment on a second hardware platform well conformed with the original results. We also selected the best augmented dataset and examined GradCAMs of several random examples from the training cohort to study effect of augmentation at the end of training phase. It was concluded that channel-independent augmentation works better in DW-MRI image for prostate. With further investigation, this may be shown to be the case for other cancer sites. We also observed that effective augmentation can help the CNN classify an augmented sample correctly while the original sample was misclassified.

We tried different augmentation combinations, however, we did not find them effective and AUC decreased. For example, we combined random rotation [-50, 50] and random crop (crop size=95) which were the most efficient candidates to the shallow CNN. We obtained AUCs of 87.86% and 82.40% on validation and test sets, respectively, which are both lower than that of using only random rotation as augmentation. Similar results were achieved for the deep CNN. With rotation, we also applied background perturbation using a mask to maintain the tumor area unchanged and only rotate background randomly to augment training dataset. The AUC results on validation and test sets set were 0.74% and 2% lower compared to rotation only.

Bae *et al.* used Perlin noise as an augmentation method to train CNN for 2D high resolution CT in diffuse interstitial lung disease images classification [24]. In their work, the accuracy with data augmentation using Perlin noise (89.5%) was significantly higher than that with conventional data augmentation (82.1%) [24]. We tried the same Perlin noise augmentation method, but the effects were not significant and the AUC results on validation and test sets improved by only 0.54% and 0.76%, respectively. Zhong *et al.* proposed Random Erasing method by cutting out an arbitrary region of the input image during each training iteration [25]. We also tried random erasing as a data augmentation method to improve performance of CNN. However, this improved the AUC only by 1.19% and 0.6% for validation and test sets, respectively.

## 5- Conclusion

Data augmentation is a useful technique for constructing enough amount of data in the limited data domain. In this paper, we proposed a fully automated method for prostate cancer detection in DW-MRI images using CNNs. We applied 5 different data augmentation methods to the training dataset of DW-MRI images of prostate. Two different deep learning models (CNNs) were trained to classify prostate cancer. Random rotation and translation were found to be the most efficient data augmentation methods for the shallow CNN and the deep CNN in prostate cancer detection, respectively.

## 6- Acknowledgment

This study received funding support in part by the Ontario Institute for Cancer Research and China Scholarship Council.

## References

[1]     H. G. Welch and W. C. Black, "Overdiagnosis in cancer," *J. Natl. Cancer Inst.*, vol. 102, no. 9, pp. 605–613, 2010, doi: 10.1093/jnci/djq099.

[2]     J. E. Thompson *et al.*, "The diagnostic performance of multiparametric magnetic resonance imaging to detect significant prostate cancer," *J. Urol.*, vol. 195, no. 5, pp. 1428–1435, 2016, doi: 10.1016/j.juro.2015.10.140.

[3]     M. I. Razzak, S. Naz, and A. Zaib, "Deep learning for medical image processing: Overview, challenges and the future," *Lect. Notes Comput. Vis. Biomech.*, vol. 26, pp. 323–350, 2018, doi: 10.1007/978-3-319-65981-7_12.

[4]     R. Cao *et al.*, "Joint Prostate Cancer Detection and Gleason Score Prediction in mp-MRI via FocalNet," *IEEE Trans. Med. Imaging*, vol. PP, no. 8, pp. 1–1, 2019, doi: 10.1109/tmi.2019.2901928.

[5]     G. Campanella *et al.*, "Clinical-grade computational pathology using weakly supervised deep

learning on whole slide images," *Nat. Med.*, vol. 25, no. 8, pp. 1301–1309, 2019, doi: 10.1038/s41591-019-0508-1.

[6] J. Ishioka *et al.*, "Computer-aided diagnosis of prostate cancer on magnetic resonance imaging using a convolutional neural network algorithm," *BJU Int.*, vol. 122, no. 3, pp. 411–417, Sep. 2018, doi: 10.1111/bju.14397.

[7] X. Wang *et al.*, "Searching for prostate cancer by fully automated magnetic resonance imaging classification: Deep learning versus non-deep learning," *Sci. Rep.*, vol. 7, no. 1, pp. 1–8, 2017, doi: 10.1038/s41598-017-15720-y.

[8] S. Liu, H. Zheng, Y. Feng, and W. Li, "Prostate cancer diagnosis using deep learning with 3D multiparametric MRI," *Med. Imaging 2017 Comput. Diagnosis*, vol. 10134, p. 1013428, 2017, doi: 10.1117/12.2277121.

[9] A. Mehrtash *et al.*, "Classification of clinical significance of MRI prostate findings using 3D convolutional neural networks," *Med. Imaging 2017 Comput. Diagnosis*, vol. 10134, no. March 2017, p. 101342A, 2017, doi: 10.1117/12.2277123.

[10] A. Esteva *et al.*, "Dermatologist-level classification of skin cancer with deep neural networks," *Nature*, vol. 542, no. 7639, pp. 115–118, 2017, doi: 10.1038/nature21056.

[11] J. Ding, B. Chen, H. Liu, and M. Huang, "Convolutional Neural Network with Data Augmentation for SAR Target Recognition," *IEEE Geosci. Remote Sens. Lett.*, vol. 13, no. 3, pp. 364–368, 2016, doi: 10.1109/LGRS.2015.2513754.

[12] J. J. Lv, X. H. Shao, J. S. Huang, X. D. Zhou, and X. Zhou, "Data augmentation for face recognition," *Neurocomputing*, vol. 230, no. December 2016, pp. 184–196, 2017, doi: 10.1016/j.neucom.2016.12.025.

[13] S. H. Park, J. M. Goo, and C. H. Jo, "Receiver operating characteristic (ROC) curve: Practical review for radiologists," *Korean J. Radiol.*, vol. 5, no. 1, pp. 11–18, 2004, doi: 10.3348/kjr.2004.5.1.11.

[14] R. R. Selvaraju, M. Cogswell, A. Das, R. Vedantam, D. Parikh, and D. Batra, "Grad-CAM: Visual Explanations from Deep Networks via Gradient-Based Localization," *Int. J. Comput. Vis.*, vol. 128, no. 2, pp. 336–359, 2020, doi: 10.1007/s11263-019-01228-7.

[15] J. A. Parker, R. V. Kenyon, and D. E. Troxel, "Comparison of Interpolating Methods for Image Resampling," *IEEE Trans. Med. Imaging*, vol. 2, no. 1, pp. 31–39, 1983, doi: 10.1109/TMI.1983.4307610.

[16] J. Glaister, A. Cameron, A. Wong, and M. A. Haider, "Quantitative investigative analysis of tumour separability in the prostate gland using ultra-high b-value computed diffusion imaging," *Proc. Annu. Int. Conf. IEEE Eng. Med. Biol. Soc. EMBS*, pp. 420–423, 2012, doi: 10.1109/EMBC.2012.6345957.

[17] K. Namdar, I. Gujrathi, M. A. Haider, and F. Khalvati, "Evolution-based Fine-tuning of CNNs for Prostate Cancer Detection," in *International Conference on Neural Information Systems (NeurIPS)*, 2019.

[18] K. Simonyan and A. Zisserman, "Very Deep Convolutional Networks for Large-Scale Image Recognition," in *International Conference on Learning Representations. (ICLR*, 2015, pp. 1–14, doi: 10.1016/j.infsof.2008.09.005.

[19]    S. Ruder, "An overview of gradient descent optimization," *arXiv*, pp. 1–14, 2016.

[20]    X. Glorot and Y. Bengio, "Understanding the difficulty of training deep feedforward neural networks," *J. Mach. Learn. Res.*, vol. 9, pp. 249–256, 2010.

[21]    S. Vaishali, K. K. Rao, and G. V. S. Rao, "A review on noise reduction methods for brain MRI images," *Int. Conf. Signal Process. Commun. Eng. Syst. - Proc. SPACES 2015, Assoc. with IEEE*, pp. 363–365, 2015, doi: 10.1109/SPACES.2015.7058284.

[22]    M. A. Islam*, S. Jia*, and N. D. B. Bruce, "How much Position Information Do Convolutional Neural Networks Encode?," in *International Conference on Learning Representations*, 2020.

[23]    S. Yoo, I. Gujrathi, M. A. Haider, and F. Khalvati, "Prostate Cancer Detection using Deep Convolutional Neural Networks," *Sci. Rep.*, vol. 9, no. 1, pp. 1–10, 2019, doi: 10.1038/s41598-019-55972-4.

[24]    H. J. Bae *et al.*, "A Perlin Noise-Based Augmentation Strategy for Deep Learning with Small Data Samples of HRCT Images," *Sci. Rep.*, vol. 8, no. 1, pp. 1–7, 2018, doi: 10.1038/s41598-018-36047-2.

[25]    Z. Zhong, L. Zheng, G. Kang, S. Li, and Y. Yang, "Random Erasing Data Augmentation," 2017.